\documentclass[fleqn,twoside]{article}
\usepackage[headings]{espcrc2}

\readRCS
$Id: espcrc2.tex,v 1.2 2004/02/24 11:22:11 spepping Exp $
\usepackage{graphicx}
\usepackage[figuresright]{rotating}

\newcommand{\AmS}{{\protect\the\textfont2
  A\kern-.1667em\lower.5ex\hbox{M}\kern-.125emS}}

\title{Iron Air collision with high density QCD}

\author{Hans-Joachim Drescher\address[MCSD]{Frankfurt Institute for Advanced Studies (FIAS),
Johann Wolfgang Goethe-Universit\"at,
Max-von-Laue-Str.~1, 60438  Frankfurt am Main, Germany}
}
       
\runtitle{Iron Air collision with high density QCD}
\runauthor{H. Drescher}

\begin{document}

\begin{abstract}
The color glass condensate approach describes successfully heavy ion
collisions at RHIC. We investigate Iron-air collisions within this
approach and compare results to event generators commonly used in air
shower simulations. We estimate uncertainties in the extrapolation to
GZK energies and discuss implications for air shower simulations. 
\vspace{1pc}
\end{abstract}

\maketitle

\section{Introduction}

Higher twist corrections to hadronic interactions become increasingly
important at high energies.  An effective way to resum all these
contributions is the color glass condensate picture.  We review the
Kharzeev-Levin-Nardi (KLN)\cite{KLN01} approach to heavy ion
collisions, which has successfully been applied to RHIC physics, and
extrapolate this model to Iron-air collisions at energies up to the
GZK ($\approx 10^{19.7}$~eV) cutoff. So far, we only calculate
multiplicities. In the near future, we hope to present a full model
which also treats the forward scattering in detail, which is important
for air showers.

\section{Review of BBL 1.0}

In Ref. \cite{Drescher:2005ig} we introduced the black body limit
(BBL) model for hadron-nucleus reactions. Valence quarks scatter
coherently off the gluon field in the target. The transverse momentum
acquired in this reaction is of the order of the saturation momentum,
defined by the density of gluons in the target. This leads to
independent fragmentation of the leading quarks at high energies, when
the saturation momentum is high. The leading baryon effect known from
lower energies is suppressed.

Gluon production in this approach is realized within the KLN model: a
simple ansatz for the unintegrated gluon distribution function (uGDF)
is applied to the $k_t$-factorization formula. In this paper, we apply
this model to Iron-air collisions. 

\section{KLN approach to nucleus-nucleus collisions}

\begin{figure}[tb]
\includegraphics[width=\columnwidth]{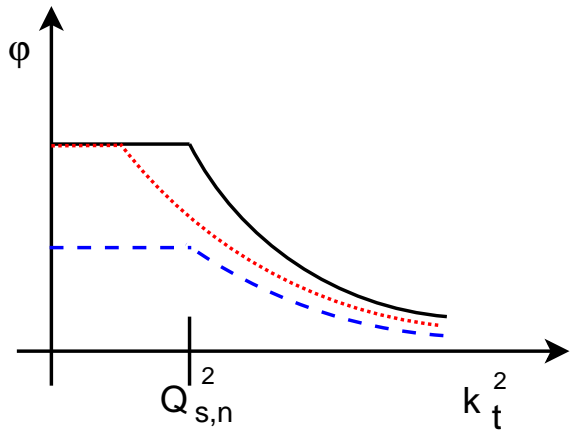}
\caption{The construction of the uGDF for low densities. The full line
shows the uGDF of a single nucleon. The dashed line is the average
uGDF for $p_A\approx0.5$. The dotted line shows the uGDF for the
averaged low density.}
\label{fig:phi}
\end{figure}

\begin{figure}[htb]
\includegraphics[width=\columnwidth]{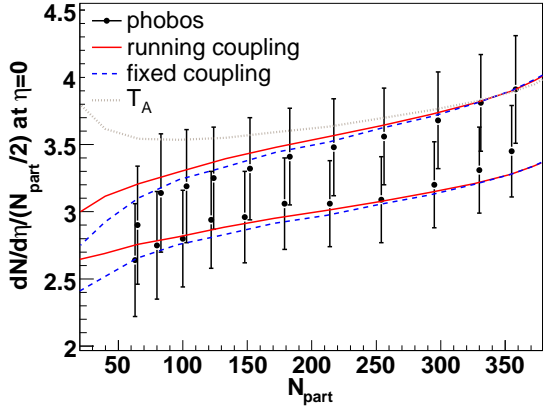}
\caption{The multiplicity of charged particles as a function of the centrality. The grey dotted line shows the result when taking $Q_s^2$ to be proportional to the average density $T_A$. The data is from PHOBOS \cite{phobos}.}
\label{fig:npart}
\end{figure}

In the $k_\perp$-factorization approach~\cite{GLR83},
the distribution of produced gluons is given by
\begin{eqnarray}
  \frac{dN_g}{d^2 r_{\perp}dy}&=&
  \frac{4N_c}{N_c^2-1} \int^{p_\perp^\mathrm{max}}\frac{d^2p_\perp}{p^2_\perp}
  \int^{p_\perp} {d^2 k_\perp} \;\alpha_s \nonumber \\
  &\times &   \phi_A(x_1, {k}_\perp^2)  \phi_B(x_2, ({p}_\perp{-}{k}_\perp)^2)
 \label{eq:ktfac}
\end{eqnarray}
with $N_c=3$ the number of colors.  Here, $p_\perp$ and $y$ denote the
transverse momentum and the rapidity of the produced gluons,
respectively. The light-cone momentum fractions of the colliding gluon
ladders are then given by $x_{1,2} = p_\perp\exp(\pm y)/\sqrt{s}$,
where $\sqrt{s}$ denotes the center of mass energy and $y$ is the
rapidity of the produced gluon.  We set
$p_\perp^\mathrm{max}$ such that the minimal saturation scale
$Q_s^\mathrm{min}(x_{1,2})$ in the above integration is
$\Lambda_{QCD}=0.2$~GeV.

The KLN approach~\cite{KLN01} employs the following uGDF:
\begin{equation}
\label{eq:uninteg}
  \phi(x,k_\perp^2;{r}_\perp)\sim
  \frac{1}{\alpha_s(Q^2_s)}\frac{Q_s^2}
   {{\rm max}(Q_s^2,k_\perp^2)}~,
\end{equation}
where $Q_s$ denotes the saturation momentum at the given momentum
fraction $x$ and transverse position ${r}_\perp$. The overall
normalization is determined by the multiplicity at mid-rapidity for the
most central collisions.  The saturation scale for nucleus $A$ is
taken to be proportional to the density of participants,
$n^A_\mathrm{part}({r}_\perp)$. This is not a universal quantity
which depends only on the properties of a single nucleus, in other
words, the uGDF is not fully factorizable.

\begin{figure}[t]
\includegraphics[width=\columnwidth]{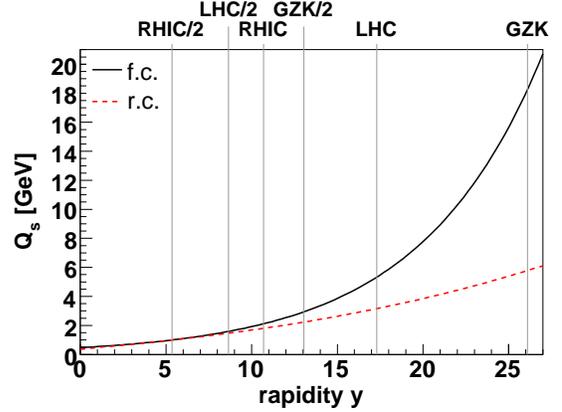}
\caption{The saturation scale as a function of the rapidity
$y=\log(1/x)$. The vertical lines denote the forward and mid-rapidities
for the different energy regions.}
\label{fig:Qss}
\end{figure}

A possible solution seems to be to define the saturation momentum
squared to be proportional to $T_A(r_\perp)$, which is dependent on
one nucleus only and therefore respects factorization. However, this
ansatz cannot describe the data on the multiplicity as a function of
centrality for Au-Au collisions at RHIC, as shown in
Fig.~\ref{fig:npart}. Why this does not work can best be seen with the
following example. We consider a peripheral Au-Au collision, and want
to construct the uGDF at the edge of one of the nuclei. Here, $T_A$ is
very small, which means that only in some nucleus configurations, we
actually find a nucleon at this position. Let us denote the
probability to find (at least) one nucleon with $p_A$. The uGDF at this
position is then $p_A$ times the uGDF of a single nucleon, as sketched
in Fig. \ref{fig:phi} (dashed line). Taking the average density $T_A$
would lead to the dotted line, and gives the wrong uGDF, since
averaging has to be done after constructing the wave function and not
before. The density under the condition to find at least one nucleon
is $T_A/p_A$, see \cite{Drescher:2006ca} for details. The saturation
scale is therefore:

\begin{equation}
  Q^2_{s,A}(x,{r}_\perp) = 
  2\,{\rm GeV}^2\left(\frac{T_A({r}_\perp)}{1.53 p_A({r}_\perp)}\right)
  \left(\frac{0.01}{x}\right)^\lambda
  \label{eq:qs}
\end{equation}
and the ansatz for the uGDF is
\begin{equation}
\phi_A = p_A ~ \phi \left( \frac{T_A}{p_A} \right)~. \label{eq:phi_factorized}
\end{equation}
In Ref.~\cite{Drescher:2006ca} we explain that the multiplicity is a
homogeneous function of order one in the density of both nuclei, and
the $n_{part}$ description is a good approximation of the factorized KLN
approach.

\section{Results}
\begin{figure}[tb]
\includegraphics[width=\columnwidth]{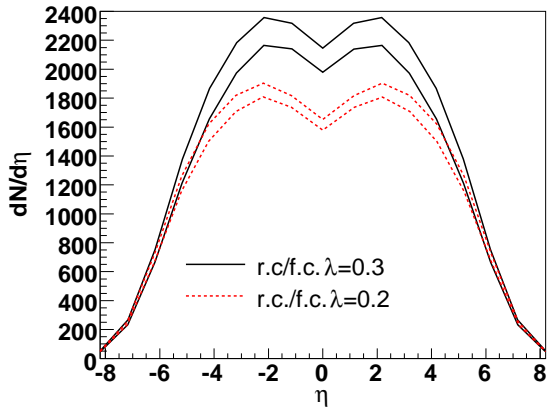}
\caption{The rapidity dependence of the multiplicity for central
($b=2.4~$fm) Pb-Pb collisions at the LHC.}
\label{fig:dndeta_LHC}
\end{figure}

\begin{figure}[tb]
\includegraphics[width=\columnwidth]{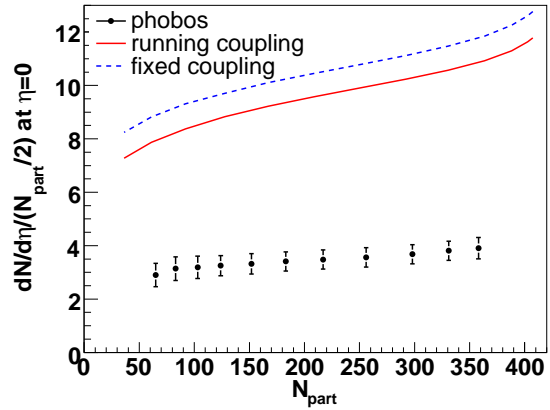}
\caption{The multiplicity of charged particles as a function of the
centrality for a 5500 GeV Pb-Pb collision at the LHC (b=2.4~fm). The
upper curve for each $\lambda$ shows fixed coupling, the lower curve
shows running coupling evolution.}
\label{fig:npart_LHC}
\end{figure}

Before showing some results and extrapolating to high energies we want
do discuss the evolution of the saturation scale as a function of the
energy. Fig.~\ref{fig:Qss} shows $Q_s$ as a function of the rapidity
for fixed coupling and running coupling evolution. See
Ref. \cite{Drescher:2005ig} for details on these two types. The
two evolution types show large differences at forward rapidity for LHC
and GZK energies. Differences at mid-rapidity for RHIC and LHC are however
less significant. Therefore, fixed and running coupling predict very
similar results for RHIC and LHC overall multiplicities.

\subsection{Accelerator Results}

Very good agreement with the data from the PHOBOS collaboration
\cite{phobos} is shown for the multiplicity at mid-rapidity as a
function of centrality, as shown in Fig.~\ref{fig:npart}.  As already
stated, the differences between fixed and running coupling evolution
are small.

In Fig.~\ref{fig:dndeta_LHC} we see predictions for central
(b=2.4~fm) Pb-Pb collisions at the LHC ($\sqrt{s}=5500~$GeV). The 
centrality dependence of the multiplicity at mid-rapidity is shown in Fig.~\ref{fig:npart_LHC}.

\subsection{Iron Air}

\begin{figure}[htb]
\includegraphics[width=\columnwidth]{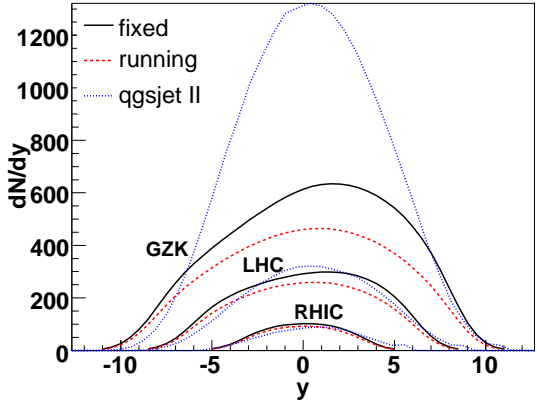}
\caption{The rapidity dependence of charged particles for central Fe-N
collisions at the three reference energies. The results for
QGSJET-II and KLN (fixed and running coupling evolution) are shown.}
\label{fig:FeN_LHC}
\end{figure}

\begin{figure}[htb]
\includegraphics[width=\columnwidth]{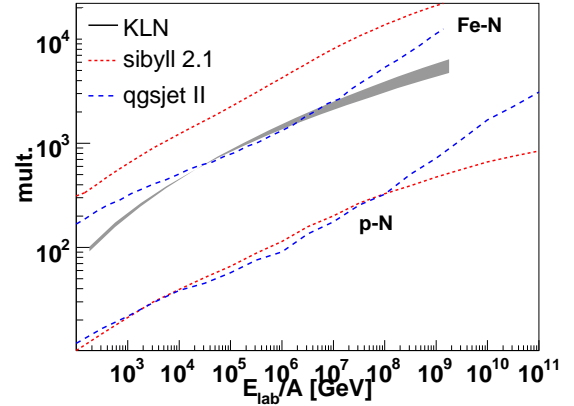}
\caption{The total multiplicity of charged particles as a function of
the energy (in the laboratory frame) per nucleon. Sibyll predicts a
higher multiplicity, since it is a superposition model and scales with
the number of participants in the Iron nucleus as compared to p-air
collision. QGSJET-II and the color glass approach predict similar
multiplicities up to LHC energies.}
\label{fig:FeN_elab2}
\end{figure}

Fig.~\ref{fig:FeN_LHC} shows the rapidity dependence of charged
particles for central Iron-Nitrogen collisions at the three reference
energies for RHC,LHC and GZK (200 GeV, 5500 GeV and 400 TeV,
respectively). Up to LHC energies, differences at mid-rapidity are
smaller than 15\%. Only at GZK energies, we see a qualitative
difference.

Fig.~\ref{fig:FeN_elab2} shows the total multiplicity of charged
particles as a function of the lab-energy per nucleon. Since in cosmic
ray physics the energy of a nucleus is usually the total energy,
plotting $E_{lab}/A$ is a useful quantity when comparing to proton-air
collisions. We also compare to the standard hadronic interactions
models Sibyll~2.1 \cite{sibyll} and QGSJET-IIc~\cite{qgsjetII}. First,
we observe that nucleus air collisions in Sibyll are obtained by
superposition of hadron air collisions. The multiplicity scales
therefore with the number of participants in the projectile and ranges
from a factor 30 to 36 compared to proton air collisions. In the
QGSJET-II model and the KLN approach, screening in the initial state
reduces the multiplicity. The predicted results of these two
approaches are quite similar up to LHC energies, whereas above this
energy, QGSJET-II predicts higher multiplicity than the color glass
approach (running coupling evolution).

\subsection{Conclusions}

We compared the multiplicities of Iron-air collisions in the color
glass condensate approach (KLN model) with hadronic interaction models
used in air shower simulations. The results of QGSJET-II and KLN are
quite similar up to LHC energies. Above this energy, QGSJET-II
predicts higher multiplicities.


\begin{thebibliography}{9}

\bibitem{KLN01}
D.~Kharzeev and M.~Nardi, 
Phys.\ Lett.\ B {\bf 507}, 121 (2001);
D.~Kharzeev, E.~Levin and M.~Nardi, 
Nucl.\ Phys.\ A {\bf 730}, 448 (2004)
[Erratum-ibid.\ A {\bf 743}, 329 (2004)];
Nucl.\ Phys.\ A {\bf 747}, 609 (2005).

\bibitem{Drescher:2005ig}
  H.~J.~Drescher,
 Nucl.Phys.Proc.Suppl. B151,163 (2006).

\bibitem{GLR83}
L.~V.~Gribov, E.~M.~Levin, and M.~G.~Ryskin,
Phys.\ Rept.\  {\bf 100}, 1 (1983).

\bibitem{Drescher:2006ca}
  H.~J.~Drescher and Y.~Nara,
  arXiv:nucl-th/0611017.

\bibitem{phobos} 
B.~B.~Back {\it et al.}  [PHOBOS Collaboration],
Phys.\ Rev.\ C {\bf 65}, 061901 (2002).

\bibitem{sibyll}
R.~S.~Fletcher, T.~K.~Gaisser, P.~Lipari and T.~Stanev,
Phys.\ Rev.\ D {\bf 50}, 5710 (1994);\\
R.~Engel, T.~K.~Gaisser, T.~Stanev and P.~Lipari,
{\it Prepared for 26th International Cosmic Ray Conference (ICRC 99),
  Salt Lake City, Utah, 17-25 Aug 1999}.

\bibitem{qgsjetII}
 S.~S.~Ostapchenko \\ Nucl.Phys.Proc.Suppl.B151(2006)143

\end{thebibliography}
\end{document}